\documentclass[prb,amsmath,amssymb,superscriptaddress,12pt,a4paper,twoside]{revtex4}

\usepackage{graphicx}

\begin{document}

\title{Turnstile behaviour of the Cooper-pair pump}


\author{J.~J.~Toppari}
\email[E-mail:]{jussi.toppari@phys.jyu.fi}
\affiliation{NanoScience Center, Department of Physics, University of Jyv\"askyl\"a,
P.O.~Box 35 (YFL), FIN-40014 University of Jyv\"askyl\"a, FINLAND}
\affiliation{Low Temperature Laboratory, P.O.~Box 2200,
FIN-02015 Helsinki University of Technology, FINLAND}
\author{ J.~M.~Kivioja} 
\affiliation{Low Temperature Laboratory, P.O.~Box 2200,
FIN-02015 Helsinki University of Technology, FINLAND}
\author{J.~P.~Pekola}
\affiliation{Low Temperature Laboratory, P.O.~Box 2200,
FIN-02015 Helsinki University of Technology, FINLAND}
\author{M.~T.~Savolainen}
\affiliation{NanoScience Center, Department of Physics, University of Jyv\"askyl\"a,
P.O.~Box 35 (YFL), FIN-40014 University of Jyv\"askyl\"a, FINLAND}

\begin{abstract}
\linespread{0.9}
\vspace{3mm}\sl
We have experimentally studied the behaviour of the so-called Cooper pair pump 
(CPP) with three Josephson junctions, in the limit of small Josephson 
coupling $E_{\mathrm{J}}\!<\!E_{\mathrm{C}}$. These experiments 
show that the CPP can be operated as a traditional turnstile device yielding a 
gate-induced current $2ef$ in the direction of the bias voltage, by applying 
an RF-signal with frequency $f$ to the two gates in phase, 
while residing at the degeneracy node of the gate plane. 
Accuracy of the CPP during this kind of operation was about 3\% and the 
fundamental Landau-Zener limit was observed to lie above 20 MHz. 
We have also measured the current pumped through the array by 
rotating around the degeneracy node in the gate plane. We show that 
this reproduces the turnstile-kind of behavior. To overcome 
the contradiction between the obtained $e$-periodic
DC-modulation and a pure $2e$-behaviour in the RF-measurements, we base
our observations on a general principle that the system always minimises its 
energy. It suggests that if the excess quasiparticles 
in the system have a freedom to tunnel, they will organize themselves to 
the configuration yielding the highest current. 
\end{abstract}


\maketitle


\section{Introduction}

During the last two decades lots of studies, both theoretical and experimental, 
have been carried out concerning the parametric pumping of charge, an idea 
originally introduced by Thouless in 1982.\cite{thou83} This phenomenon is 
based on the ability of a propagating potential well to carry a charge $q$ 
through a system.  This again makes possible controlled pumping, 
by periodically changing the system parameters at a 
frequency $f$ to induce propagation of charge during every 
cycle. This yields a DC-current $I\!=\!qf$ through the system.

The parametric pumping of charge can be obtained in many different 
kinds of devices, but most of the attention has been directed towards
the systems where the charge $q$ passed through during each cycle
is quantised at a certain number of electrons, $q\!=\!-ne$, where $n$ is an
integer and the elementary charge of an electron $e\!=\!\!|e|$ is defined as 
a positive number through the paper. This can be realised, e.g., by semiconductor 
quantum 
dots by varying the height of the tunnelling barriers\cite{kouw91} 
or by one-dimensional ballistic channels in a so-called SAW-pump, where the 
transport is induced by an acoustoelectric wave (SAW).\cite{shi96,tal97}
The most promising candidate so far is the so-called single electron pump
consisting of an array of three or more mesoscopic metallic
tunnel junctions in the Coulomb blockade regime.\cite{pot91,kel96,kel98} 
Due to  Coulomb blockade the number of electrons in the islands of 
the array is very accurately controlled and the pumping can be induced 
by phase-shifted gate voltages,\cite{pot91,kel96,kel98} which yield a current 
$I\!=\!-nef$. Here $f$ is the frequency of the RF-signal applied to the gates and 
the integer number $n$ depends on the amplitude of the 
operating trajectory in the gate variables. 
These devices are more accurate than those based on semiconductors, 
even sufficient for metrological 
applications.\cite{kel96,kel98} In the past the 
single electron pump has been proposed
to close the metrological triangle by providing the standard for 
electrical current.\cite{lik85,kel96,kel98} 

The only drawback in the single electron
pump is the low operating frequency $f\!\alt\!5$ MHz, which cannot provide
high enough current for a current standard. At higher frequencies 
the accuracy is lost due to coherent higher order charge
transfer processes known as (in)elastic co-tunneling\cite{ave90,ave89,geer90} 
and other sources of error discussed in 
Refs.~\onlinecite{jen92,ave93,mart94,fon96}. 
The maximum operating frequency could 
possibly be pushed higher by using a Cooper pair pump (CPP) 
consisting of  three or more small Josephson junctions in series. The pumping
in the CPP is achieved similarly by gate voltages and the charge is quantised 
at discrete numbers of Cooper pairs in the islands yielding a current
$I\!=\!-n2ef$. Due to the different nature of 
Cooper pair tunnelling the operating frequency is now limited by the 
Landau-Zener (LZ) transitions,\cite{ziman} which yields several 100 MHz for the upper
limit of the operation frequency, depending on the parameters of the device.   
Due to the coherent nature of Cooper pair tunnelling, the CPP is also subject 
to intensive cotunnelling, which reduces the accuracy significantly 
in short arrays.\cite{pek99,aun00} To increase the accuracy one should
use longer arrays or suppress the cotunnelling by other means.\cite{zorpump,sluice}

Another reason for interest in Cooper pair pumping is quantum computing, 
where Josephson junction circuits look promising due to their easy scalability 
and relatively long decoherence times.\cite{nak99,fried00,vanderwal00,vion02,mak01}
It has been proposed that two capacitively coupled Josephson junction arrays
could form a quantum bit (qubit) and CPP could also be used to transport
the information in a more complicated device.\cite{ave98} When operated
in suitable electromagnetic environment the CPP could also provide a method 
to directly measure the decoherence rate.\cite{pek01,faz03}    

In this article we concentrate on transport properties of the CPP and 
how the non-idealities, e.g., dissipation, quasiparticle tunnelling and 
strong cotunnelling affect its performance.

\section{Cooper pair pump}
\label{sec:model}

The Cooper pair pump (CPP) consists of three or more mesoscopic Josephson junctions in 
series with gate voltages capacitively coupled to each of the islands
between the junctions. It is characterized by several important energy scales. 
The first one is the typical Josephson coupling energy
$E_{\rm J}$ of Cooper pair tunnelling through the junctions
and the second one is the charging energy $E_{\rm C}$ due to 
the small islands between the junctions. 
The ratio of these two, $E_\mathrm{J}/E_\mathrm{C}$, is an important 
parameter in determining the  
behaviour of the device. To be able to obtain pumping of single charges,
we have to restrict ourselves to
the limit $E_\mathrm{J} \ll E_\mathrm{C}$, where the dynamics are mainly
determined by discrete tunnelling of charge carriers, i.e., Cooper pairs.
In addition, the charging energy has to be larger than the thermal energy 
$k_\mathrm{B}T$ to prevent thermal excitations, but smaller than
the superconducting gap $\Delta$ in the density of quasiparticle states
to prevent quasiparticle poisoning.\cite{mat93,joy94} This yields the usual chain
of inequalities $ k_\mathrm{B}T< E_\mathrm{J}\ll E_\mathrm{C}<\Delta$. 

\begin{figure}[t]
\linespread{1}
\includegraphics[width=110truemm]{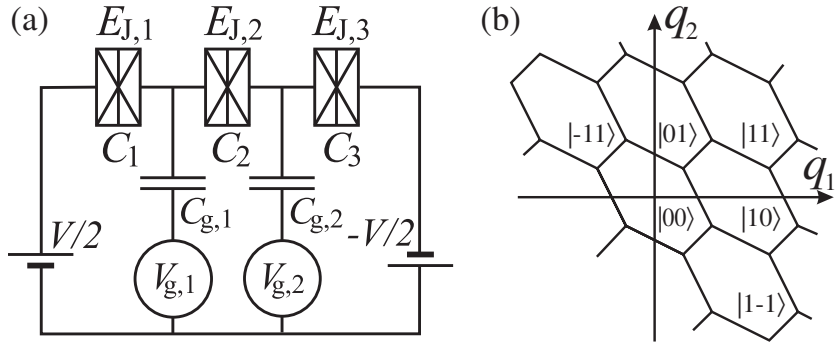}
\caption{(a) A superconducting array of three Josephson junctions (CPP). 
Here $C_k$ and $E_{\mathrm{J},k}$ are the capacitance and the Josephson 
energy of the $k$th junction, respectively. (b) Stability diagram of a uniform CPP
at zero bias $V=0$ on the plane determined by the normalised gate 
charges $q_i = V_{\mathrm{g},i} 
C_{\mathrm{g},i}/2e$. The stable configuration inside each of the hexagons 
is shown by the kets $\vert n_1n_2\rangle$. 
\label{fig:pump}}
\end{figure}

In this article we consider a three junction CPP whose circuit
schematics are shown in Fig.~\ref{fig:pump}(a).
Each  junction has a capacitance $C_k$ and a Josephson energy 
$E_{{\rm J},k}$. Gate voltages $V_{\mathrm{g},k}$ are 
assumed to be externally operated and coupled to islands with capacitances
$C_{\mathrm{g},k}$. The Josephson phase 
difference $\varphi$ across the array is related to the bias voltage 
$V$ according to the relation $d\varphi/dt=(-2e)V/\hbar$.

Including only Josephson and charging energies and
neglecting quasiparticle tunnelling as well as other degrees of freedom,
the Hamiltonian of the system can be written as\cite{geer,pek99} 
\begin{equation}
H=H_{\mathrm{Ch}}(q_1,q_2)+H_{\mathrm{J}},
\label{eq:hamilton}
\end{equation}
where the charging Hamiltonian 
$H_{\mathrm{Ch}}(q_1,q_2)$ depends on the normalised gate charges,
$q_i = V_{\mathrm{g},i} C_{\mathrm{g},i}/2e$,  and 
the  number of Cooper pairs on each island, $n_i$, according to
$\langle n_1n_2\vert H_\mathrm{Ch}(q_1,q_2)\vert n_1'n_2'\rangle_{\varphi}=
E_\mathrm{Ch}(u_1,u_2)\delta_{n_1,n_1'}\delta_{n_2,n_2'}$, 
where $u_i = n_i - q_i$. 
This model Hamiltonian is described in detail in Refs.~\onlinecite{aun00}
and \onlinecite{aun01}. The function $E_\mathrm{Ch}(u_1,u_2)$ gives the 
details of the charging energy and in the presence of the bias voltage 
$V$ it attains a form
\begin{equation}
E_\mathrm{Ch}=\frac{4E_\mathrm{C}\lambda_\mathrm{g}}
{\lambda_\mathrm{g}^2-1}\left[u_1^2+u_2^2 + 
\frac{2}{\lambda_\mathrm{g}}u_1u_2
-\frac{CV}{e}\left(\frac{u_1}{\lambda_\mathrm{g}} 
+ u_2\right)\right]-2peV,
\label{eq:ech}
\end{equation}
where we have assumed a symmetric array with $C_1\!=\!C_2\!=C_3\!\equiv\!C$ and
$C_{\mathrm{g},1}\!=C_{\mathrm{g},2}\!\equiv\!C_\mathrm{g}$. Parameter
$\lambda_\mathrm{g} = 2 + C_\mathrm{g}/C$, $E_\mathrm{C}=e^2/2C$ 
is the unit of charging energy, and $p$ is the number 
of Cooper pairs tunneled through the whole array. 
The Josephson (tunnelling) Hamiltonian is given by 
\begin{equation}
H_{\mathrm{J}}=-\sum_{k=1}^3 E_{\mathrm{J},k}\cos(\varphi_k),
\label{eq:tunnelit}
\end{equation}
where $E_{\mathrm{J},k}$ and $\varphi_k$ are the Josephson coupling energy
and the phase difference across the $k$th junction, respectively. 

At zero bias, $V=0$,  Eq.~(\ref{eq:ech}) yields a honeycomb like stability diagram 
shown
in Fig.~\ref{fig:pump}(b).\cite{geer,pek99} Inside each hexagon the system is stable 
and there is 
one charge state $\vert n_1n_2\rangle$, i.e., the eigenstate of the 
charging Hamiltonian, $H_\mathrm{Ch}$, as a ground state. At the edges of
hexagons  and at the triple nodes, two or three charge states 
are degenerate, respectively. The effect of the Josephson coupling, 
$H_\mathrm{J}$, is to eliminate these degeneracies by introducing
a coupling between different charge states. This induces new 
eigenstates which are superpositions of charge states and an energy gap 
will open between different eigenstates near the degeneracy lines of the
charging part of the energy.\cite{pek99, zorset}
 
\begin{figure}[b]
\linespread{1}
\includegraphics[width=80truemm]{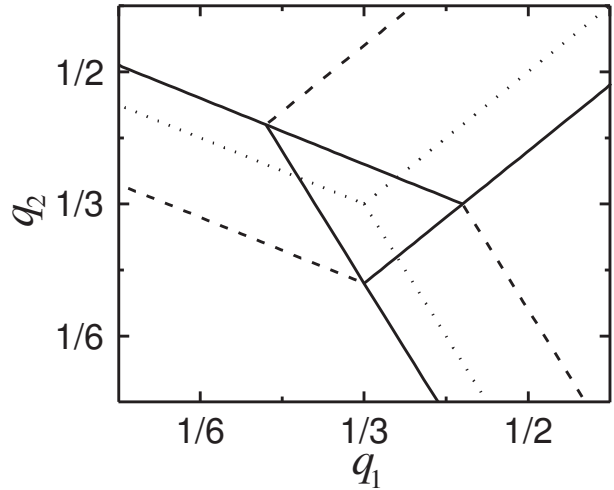}~~~
\caption{Zoomed view of one of the nodes in the stability diagram of the CPP 
(see Fig.~\ref{fig:pump}(b)). The dotted lines correspond to $V=0$. 
Solid and dashed lines show the resonance condition for
Cooper pair tunnelling and cotunnelling in the presence of bias voltage
$CV/2e = 0.1$. 
\label{fig:triangle}}
\end{figure}

Tunnelling events in CPP take place as a coherent tunnelling of Cooper pairs, 
in which the system travels adiabatically from the initial charge state to the 
final one along the eigenstate of the full Hamiltonian of Eq.~(\ref{eq:hamilton}).
This eigenstate is a superposition of the charge states with coefficients varying 
continuously when changing the gate charges, thus resulting in 
a tunnelling of a Cooper pair when a resonance in $(q_1,q_2)$ plane
is passed.\cite{ave91,bri91} Here the resonance
means that the initial and the final charge state have the same energy 
$E_\mathrm{Ch}$. These resonances for coherent Cooper pair 
tunnelling are shown as dotted and solid lines in Fig.~\ref{fig:triangle}
at $V=0$ and $V > 0$, respectively.  
Similar resonances for the second order process, cotunnelling 
of a Cooper pair, are shown by dashed lines. 
By cotunneling we mean the coherent tunnelling through 
two junctions simultaneously which is qualitatively similar 
to the cotunnelling in normal state.\cite{ave90,ave89,geer90} Still higher order
processes are weak and play insignificant role with our sample parameters.
These resonances of coherent tunnelling and cotunnelling overlap in 
case of $V=0$, but separate as non-zero bias voltage $V$ is applied.
Thus, each degeneracy node is split into a triangle
determined by the first order tunnelling resonances,
as shown by solid lines in Fig.~\ref{fig:triangle}.\cite{geer,bithesis}
Inside the triangles  the state of the CPP  
depends on the path along which the system has reached the point, 
thus opening the possibility for hysteretic behavior.

Pumping of Cooper pairs at $V=0$ is achieved by adiabatically  varying the gate 
voltages along the path encircling one of the degeneracy nodes. 
The principle of pumping Cooper pairs is explained, e.g., in Refs.~\onlinecite{geer},
\onlinecite{pek99} and \onlinecite{aun00}.  The obtained adiabatic
evolution of the eigenstates splits the transferred charge into two
parts: \cite{pek99,aun00} The pumped charge $Q_\mathrm{P}$ 
and the charge $Q_\mathrm{S}$ carried by the constantly flowing supercurrent 
$I_\mathrm{S}$. The latter one of these can be calculated as the $\varphi$-derivative 
of the
dynamical phase $\eta_m=-\int_0^t (E_m(\tau)/\hbar)d\tau$ in the state
$\vert m \rangle$, $Q_\mathrm{S}/2e=-\partial\eta_m/\partial\varphi$,
while the pumped charge is related to the $\varphi$-derivative of Berry's 
phase\cite{ber84} 
$\gamma_m=i\oint \langle m\vert dm\rangle$ attained 
along the pumping path, $Q_\mathrm{P}/2e=-\partial\gamma_m
/\partial\varphi$.\cite{pumpberry}

\section{Sample fabrication}

\begin{figure}[tb]
\linespread{1}
\includegraphics[width=82truemm]{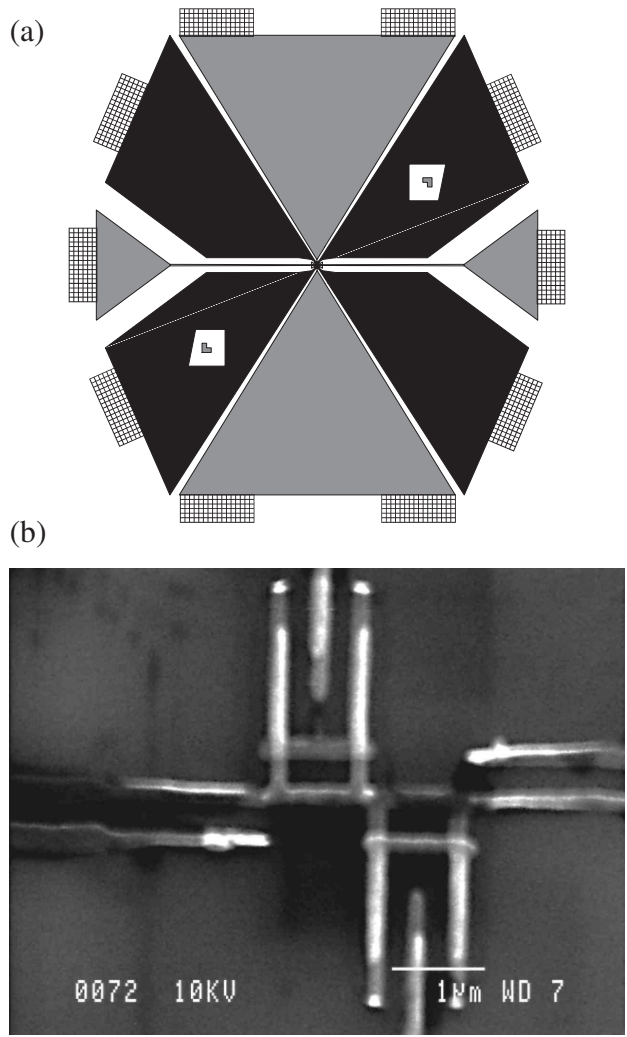}
\caption{(a) Scheme of the large scale structures of the sample. 
Gray areas indicate biasing lines and the triangular ones gate 
lines at the top and the bottom. Black areas show the large guard planes to suppress
parasitic crosscouplings. Grid-like areas in the contact pads are used in 
ultrasonic bonding, and they improve the mechanical strength of the contact. 
Also the two bigger alignment marks are seen inside the two guard planes.
(b) An SEM-image of the small structures of the sample used in these experiments. 
The interdigital type design of the gates lowers the crosscoupling 
between them. The large guard planes as well as quasiparticle traps fall 
outside the image. 
\label{fig:SEM}}
\end{figure}

The sample used in experiments was fabricated by e-beam lithography with a 
conventional 
self-aligning shadow angle evaporation technique. The two different
{\em ex situ} process steps were performed to make it possible to first 
evaporate large scale structures containing all the contact pads for bias and 
gate lines and the large guard planes isolating these various entries. 
These guard planes were permanently grounded by 
bonding them to the bottom of the sample holder.
The undesirable capacitive crosscouplings between the lines
were eliminated this way effectively as will be seen later. The design
of the large scale structures is shown in Fig.~\ref{fig:SEM}(a).
These large structures were fabricated of gold to ensure good electrical
contact in ultrasonic bonding of wires on the sample stage,
and between the two layers fabricated {\em ex situ}. The 
so-called quasiparticle traps in biasing lines near the sample 
were also evaporated at the same time.\cite{laf93,joy94}
These quasiparticle traps consisting of a 2.5 
$\mu$m $\times$ 0.6 $\mu$m $\times$ 60 nm sheet of gold were placed 
under the biasing line about 4 $\mu$m away from the outermost junctions. 

The sample itself is surrounded by these large scale structures and it 
consists of three Josephson junctions of about 100 nm $\times$ 100 nm
in area. The two $\sim 1$ $\mu$m long islands had an interdigital 
type of design to improve the gate coupling and to suppress 
the crosscouplings. The gate lines were also brought to the sample via 
a 1.8 $\mu$m wide channel between the guard planes. The small 
structures were fabricated out of aluminium with aluminium oxide as a
barrier in the tunnel  junctions. 

\section{Characterisation of the sample}

\subsection{Experimental setup}

\begin{figure}[t]
\linespread{1}
\includegraphics[width=80truemm]{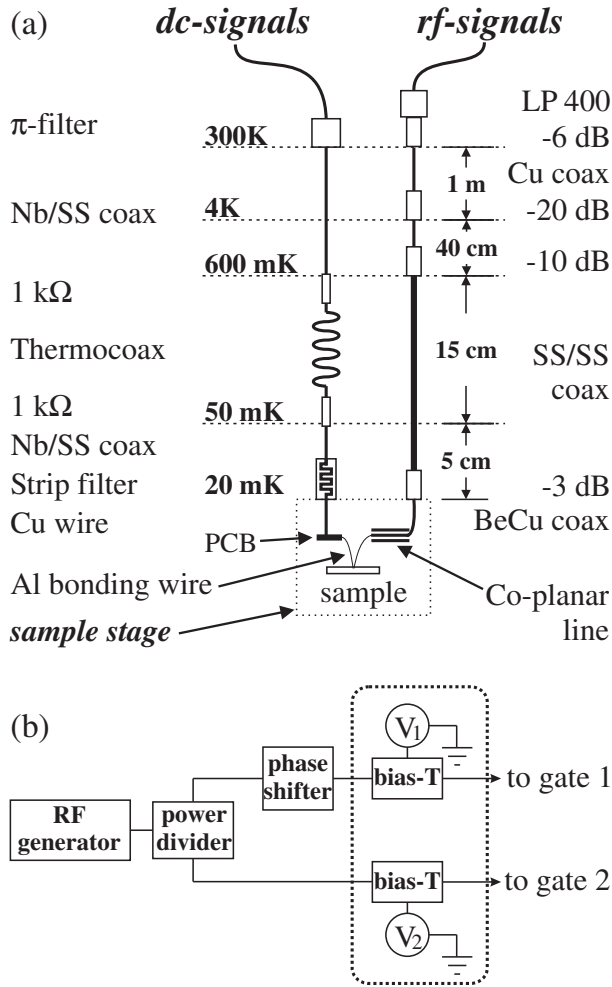}
\caption{\label{expfig} 
(a) Measurement wiring of the dilution refrigerator. (b) 
A schematic picture of the room temperature RF-signal connections. The
dotted line shows the part included in the integrated domestic circuit. 
The compensation circuit was not needed in the measurements due to
low crosscoupling in the sample. }
\end{figure}

The experimental setup is shown in Fig.~\ref{expfig}(a).
All measurements were done in an S.H.E. Corporation DRI-420 
dilution refrigerator whose minimum temperature is $\sim 10$ mK.
Lower parts of the cryostat are surrounded by lead over the vacuum 
jacket in the helium bath for magnetic shielding.
 
The fridge has 14 highly filtered lines for DC-signals (Fig.~\ref{expfig} (a)). These 
lines include
3 stage low pass filtering at different temperatures. At room temperature we used
commercial low pass $\mathrm{\pi}$-filters (Tusonix 4101, -55 dB at 100 MHz) which 
were 
connected directly to the top of the cryostat. From room temperature down to 600 mK 
all DC-signals 
are fed through coaxial cables with Nb as an inner conductor and stainless steel as 
shielding. Between 600 mK 
and 60 mK plate each line has 1.5 m of Thermocoax$^{\textcircled{\scriptsize
R}}$ cable which also forms the next filtering stage (-200 dB at 20 
GHz).\cite{thermocoax}  
At both ends of Thermocoax$^{\textcircled{\scriptsize
R}}$ cables there are 1 k$\Omega$ resistors in series to improve filtering 
at low frequencies ($< 1$ GHz). The last filtering stage is at the sample 
stage at the base temperature. These filters were 
commercial stress gauges (KYOWA KFG-2-350-D1-23) squeezed between two ground planes 
forming
continuous RC strip line filters.\cite{strip} Short sample stage wires 
were made of Cu and were soldered to a printed circuit board (PCB). This again was 
ultrasonically bonded to the DC electrode of the sample with aluminum wires.

Our refrigerator has 4 lines for RF-signals (Fig.~\ref{expfig} (a)). At room 
temperature
we used 400 MHz low pass filters (Mini-circuits SBLP-400) and -6 dB fixed attenuators 
(Inmet). 
These were directly connected to the top of the cryostat and 
all other room temperature connections were made by
using flexible SUHNER Sucoflex 104P cables with SMA connectors. From room temperature 
down to 4 K RF-signals are fed through Cu coaxial cables. At low temperatures we use 
BeCu coaxial
cables, except between 600 mK and the sample stage we use semirigid stainless 
steel coaxial cables for better thermal isolation. As a whole the RF lines have -33 dB 
attenuation
at low temperatures: -20 dB at 4 K, -10 dB at 600 mK and -3 dB at sample stage 
temperature (Inmet fixed attenuators). 
At the sample stage we use MCX connectors and all other connectors are of SMA type. 
The sample is directly ultrasonically bonded by Al bonding wire to a coplanar 
transmission line mounted on the sample stage.
When all these lines are connected to the coldest parts of the refrigerator, 
the base temperature is lifted up to $\sim 20$ mK, as compared to the $\sim 10$
mK base temperature without these lines.

All DC-voltage measurements were done by using HMS Electronics model 568 
low-noise preamlifiers and for current measurements we used DL-Instruments 1211 
preamplifier.
Both amplifiers were powered by battery sources only. Between preamplifier and data 
aquisition
(NI PCI-6036E DAQ card) we use a home made battery powered analog optoisolation to 
avoid ground loops and digital noise in our measurements.
In Fig.~\ref{expfig}(b) we show a schematic picture of gating signal connections.
For this we used a HP8656B signal generator and divided 
the RF-signal by using an INMET 6014-2 power 
divider. One of these signals was fed through 
a phase shifter while the other went directly into the home made circuit 
schematically consisting of two bias-Ts and a 
possibility to add negative crosscoupling between the two signals
to compensate the undesirable capacitive coupling between the gates in the CPP. 
The circuit also contained a high quality RF-circuitboard and optoisolated linking to 
the 
computer which could be used to program all the gains used in bias-Ts, DC-offsets,
compensation and as a main amplification. The control program also included 
possibility 
to automate the measurements to some extent.\cite{sampo}

\subsection{Current-Voltage dependence}

\begin{figure}[htb]
\linespread{1}
\includegraphics[width=90truemm]{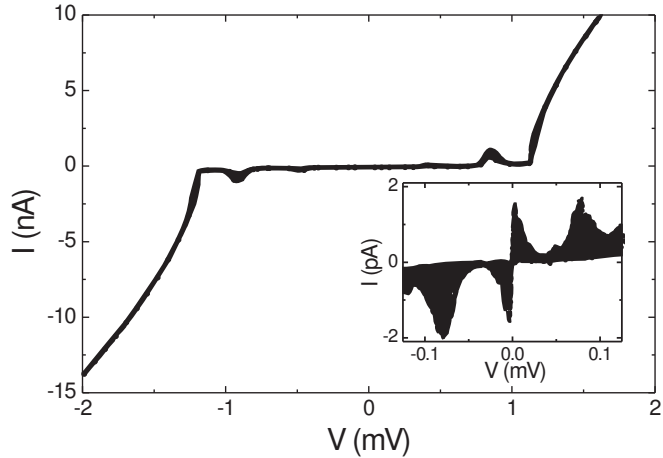}
\caption{I-V characteristics of the measured sample, taken 
at different combinations of gate voltages $V_{\mathrm{g},1}$ 
and $V_{\mathrm{g},2}$.  The inset presents the blow up of the 
I-V curves near zero bias.
\label{fig:IV}}
\end{figure}

In Fig.~\ref{fig:IV} we present I-V characteristics of the sample, taken 
with different combinations of gate voltages $V_{\mathrm{g},1}$ 
and $V_{\mathrm{g},2}$. It shows a sharp rise at the beginning of the 
quasiparticle tunnelling  branches at bias voltages 
$V\simeq\pm 6\Delta_\mathrm{Al}/e \approx\pm 1.2$ mV. Also all four
major peaks in subgap regime corresponding to different possibilities of 
Josephson-quasiparticle (JQP) -cycles are clearly 
visible.\cite{ful89,ave89b,bri91,nak96} The gate modulation is most 
pronounced at the gap edge as well as in the region of JQP-peaks.
From the asymptotic slope of the I-V curve at high voltages we
obtain the normal state resistance per junction, $R_\mathrm{T}\!\simeq\!34$ k$\Omega$, 
which yields $E_\mathrm{J} \!=\! \Delta_\mathrm{Al}
R_\mathrm{Q}/2R_\mathrm{T} \!\approx\! 19$ 
$\mu$eV. Here we have assumed that all junctions are identical and that the critical 
current
obeys the Ambegaokar-Baratoff relation.\cite{amb63,tink96} $R_\mathrm{Q} 
\!=\! h/(2e)^2$ is the resistance quantum for Cooper pairs. The value of the charging 
energy, $E_\mathrm{C} \!\approx\! 129$ $\mu$eV, 
was obtained using the depth of the dip in the normal state conductance 
curve taken at 4.2 K.\cite{jj97} This yields 
$E_\mathrm{J}/E_\mathrm{C} \!\approx\! 0.15$. Since the charging
energy is well below the superconducting gap 
$E_\mathrm{C}\!<\!\Delta_\mathrm{Al}\approx 200$ $\mu$eV, the parity 
effect should not be suppressed.\cite{ave91,tuo92} 

The capacitances of the tunnel junctions, $C_\mathrm{T} \approx 0.62$
fF, can be calculated using the $E_\mathrm{C}$ measured as described
above, while the gate capacitances, 
$C_{\mathrm{g},i} \!\approx\! 70$ aF ($i = 1,2$), 
were obtained from the periodicity of the current modulation at high bias. 
There exists also unavoidable crosscouplings, i.e., capacitive coupling of 
the gate voltage $V_{\mathrm{g},1}$ to the second island, $C_{12}$,
and vice verse, $C_{21}$. These parasitic capacitances 
can be estimated from the slope of any known horizontal or 
vertical periodic structure in the modulation 
plane similar to the one in Fig.~\ref{fig:honey}, where the current 
is plotted as a function of the two gate voltages $V_{\mathrm{g},i}$. 
We found crosscoupling to be $C_{12}\!\approx\! C_{21}
\!\sim\! 0.16 C_\mathrm{g}$, which is low enough to make it
unnecessary to apply any active compensation for that. 

In the large scale figure the current around zero bias is not visible but it
is seen in the blow up. The supercurrent, i.e., the feature around zero bias, depends 
strongly on
the gate voltage. It is peaked around $V\!=\!0$, getting the 
maximum at $|V|\!\ll\! E_\mathrm{C}/e$ (see Fig.~\ref{fig:IV}). 
The shape of supercurrent also indicates the presence 
of the electromagnetic environment $Z(\omega)$ with low
impedance of $0 \!<\! \mathrm{Re}[Z(\omega)]\!\ll\! R_\mathrm{Q}$.\cite{ing92}
This agrees well with the impedance of the biasing lines, $\mathrm{Re}[Z(\omega)]
\!\sim\!\sqrt{\ell/c}$, which is of the order of the vacuum impedance 377 $\Omega$. 
Here $\ell$ and $c$ are the inductance and the capacitance per unit length, 
respectively, and the resistance of the normal metal sections, $R\sim\!1$ $\Omega$, 
is negligible. Resonance peaks symmetrically around zero bias at $V\!\approx\!\pm 80$ 
$\mu$V, 
are also visible in this figure, and they will be discussed later.

\subsection{Gate modulation}
\label{sec:mod}

\begin{figure}[ht]
\linespread{1}
\includegraphics[width=100truemm]{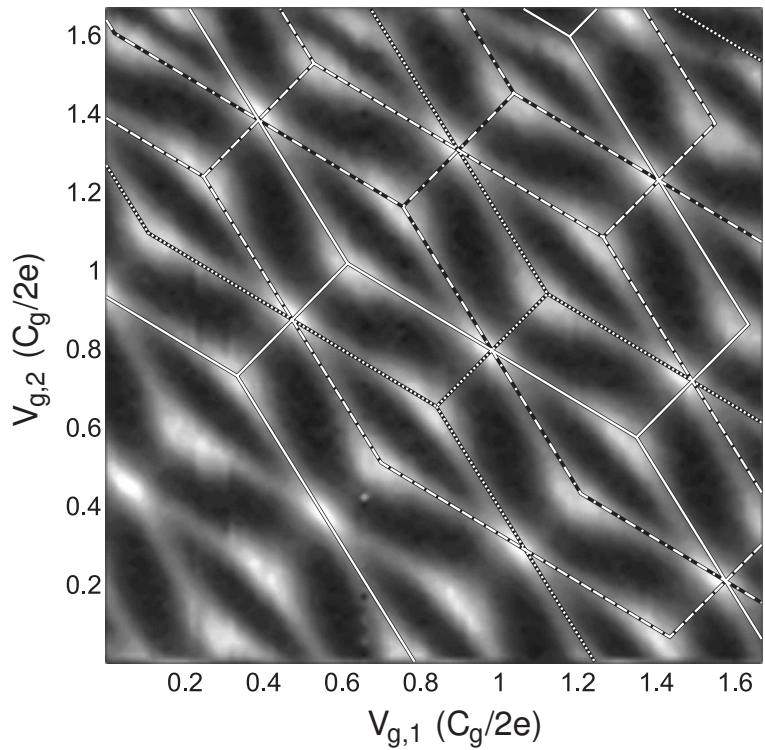}
\caption{Supercurrent $I_\mathrm{S}$ as a function of two gate voltages 
$V_{\mathrm{g},1}$~and $V_{\mathrm{g},2}$. The lighter the color is, the 
higher is the current. The structure is composed of four different 
$2e$-periodic honeycombs each corresponding to the expected 
stability diagram of the CPP. These honeycombs are separated by half 
a period due to different number of quasiparticles, which act like discrete variations 
in the gate charge. Lines drawn in the figure present each one of these 
different honeycombs in the case of a symmetric array and they fit the data exactly, 
with the gate capacitances $C_{\mathrm{g},i}$ and the crosscouplings, 
$C_{12}$ and $C_{21}$, as fitting parameters.
A small bias voltage of $V\sim 15$ $\mu$V was applied during the measurement to sit
approximately at the maximum of the supercurrent.
\label{fig:honey}}
\end{figure}

To find the correct working point for pumping we mapped out the gate 
dependence of the supercurrent $I_\mathrm{S}(V_{\mathrm{g},1},
V_{\mathrm{g},2})$ by applying a small  bias voltage to maximize 
$I_\mathrm{S}$ and measuring the current while passing systematically 
all combinations of gate voltages. 
Since $I_{\mathrm{S}}\!=\!-(2e/\hbar)(\partial H/\partial\varphi)$
it is clear that it should follow the honeycomb like stability diagram of the CPP
(See. Fig.~\ref{fig:pump}(b)), getting an increase at every 
degeneracy line and maximum value at the triply degenerate nodes.
The modulation curve obtained experimentally is presented as a 
contour plot in Fig.~\ref{fig:honey}(a). It does not show the expected
pure honeycomb like structure, which indicates
the existence of non-equilibrium quasiparticles in our system. 
Yet, the pattern cannot be ascribed to pure quasiparticle tunnelling either. 
We have observed this kind of a modulation pattern 
in most of the (more than ten) measured samples.

The effect of quasiparticles tunnelling into and out from the islands 
is to change the gate charge by one electron and thus to shift the 
$2e$-periodic stability diagram half a period (i.e., $e$-period) 
in the corresponding direction. 
If the rate of these tunnellings is higher than inverse of the measuring time, 
$\tau_m\geq 100$ $\mu$s, the current obtained is an average over many 
$2e$-periodic honeycombs separated by half a period in the direction of 
one of the two gate charges.
Since the shift due to quasiparticle tunnelling is exactly half the period of 
the stability diagram, two shifts in one direction restores the original 
honeycomb pattern. This feature can be seen in
the charging Hamiltonian, $H_\mathrm{Ch}$, i.e., tunnelling of two 
quasiparticles through the same junction has the same effect on the 
charging energy as a tunnelling of a Cooper pair. Thus, any quasiparticle 
configuration $(n_{\mathrm{qp},1},n_{\mathrm{qp},2})$ can be reduced to 
$\{\tilde n_{\mathrm{qp},i}=n_{\mathrm{qp},i} \mod 2\}_{i=1,2}$
and it is enough to consider only four different honeycombs corresponding
to, e.g., quasiparticle distributions $(0,0)$, $(1,0)$, $(0,1)$ and $(1,1)$.

As expected, the measured pattern of Fig.~\ref{fig:honey} composes of 
four shifted $2e$-periodic  honeycombs each corresponding to the stability 
diagram of the CPP in the absence of quasiparticles. 
These four honeycombs each displaced by half of the $2e$-period are illustrated 
as lines. The lines are fitted to the data with the 
gate capacitances and crosscouplings as fitting parameters and they  
correspond to the symmetric array.  
In addition to the regular degeneracy nodes the pattern shows some
extra peaks which can be explained as intersections of different 
honeycombs. In these points the supercurrent is high in three out
of four different quasiparticle configurations, yielding a
resonance peak.  

In the preceding paragraphs the measured modulation pattern consisting 
of four shifted honeycombs was explained by quasiparticle tunnelling, 
which was assumed to happen fully stochastically in both time and 
direction. This is the usual assumption in mesoscopic
superconducting devices consisting of Josephson junctions.\cite{tink95}
However, we can also explain these four honeycombs by means of quasiparticle
tunnelling and by assuming that {\it the system with degrees of freedom evolves 
via states of the lowest energy.} In case of a biased array, the system can 
always lower its energy by increasing the number $p$ of Cooper pairs 
tunneled through it. Thus, it will try to maximize the current. 
If we consider the system as a whole including quasiparticles 
and assume them to have a freedom to tunnel, they will organize themselves 
to the configuration yielding the highest current. Thus, the quasiparticle 
tunnelling events are still happening stochastically
in time but not in direction. However, 
this means that the quasiparticles do not carry the current themselves, 
which can be the case due to Coulomb blockade and the energy gap $\Delta$.  
This possibility of quasiparticles
to tunnel can be clearly observed in Fig.~\ref{fig:honey} and it can be 
justified by the same means as in the fully stochastic model.\cite{tink95}
 
According to {\it energy-minimisation} the system always changes the 
quasiparticle configuration to the one 
corresponding to maximum supercurrent, while varying the 
voltages $V_{\mathrm{g},i}$ in the measurement of the gate modulation.
Thus, this model yields the same combination of four honeycombs for the measured 
current, as we 
would obtain if the quasiparticle configurations changed fully stochastically. 
However, in our model the current is higher because it is not the time average 
over many configurations but the largest possible. Unfortunately, by 
measuring the gate modulation it is impossible to resolve 
between the fully stochastic and the energy-minimisation models and find out which 
one is the valid explanation.  Later we argue for the latter explanation 
based on the results of the RF-measurements.

However, neither of the previous models for quasiparticle tunnelling explain  
why sometimes the $2e$-periodicity is seen and sometimes not
even the sample parameters should yield the clear parity effect.
To further examine this, we can make a 
'worst case' assumption that quasiparticles always have freedom to move (due to, 
e.g., extra subgap quasiparticle states in the samples). Then the quasiparticle 
tunnelling rate $\Gamma$ would be essentially determined by the charging 
energy and the general golden rule expression (in the limit $E_\mathrm{C}\gg 
k_\mathrm{B}T$): $\Gamma \approx E_\mathrm{C}/(e^2R_\mathrm{T}^*)
\exp(-E_\mathrm{C}/k_\mathrm{B}T)$, where $R_\mathrm{T}^* = 
R_\mathrm{T}/\eta^2$ and $\eta \simeq 10^{-4}$ is the relative
density of quasiparticle states inside the gap.\cite{dyn84,pek03}
According to our sample parameters and with temperature of 30 mK, 
this yields $\sim\!10^{55}$ hours for 
the average time between the quasiparticle tunnelling events, which 
is infinite in the time scale of the measurements. Yet, this time depends strongly on
the temperature and if the electronic temperature of the sample would be higher, e.g., 
300 mK, which could be the case due to inadequate filtering or thermalisation of 
the measurement lines, the time between the quasiparticle 
tunnelling events would be $\sim$ 500 ms resulting in $e$-periodicity in the 
measurement.
This argument yields the same threshold temperature of $\sim$ 250 mK for 
$2e$-periodicity as the earlier measurements in Refs.~\onlinecite{tuo92}, 
\onlinecite{laf93} and \onlinecite{joy94}, and
could explain the lack of the $2e$-periodicity in some cases.

\subsection{Effect of the bias voltage}

\begin{figure}[bht]
\linespread{1}
\includegraphics[width=87truemm]{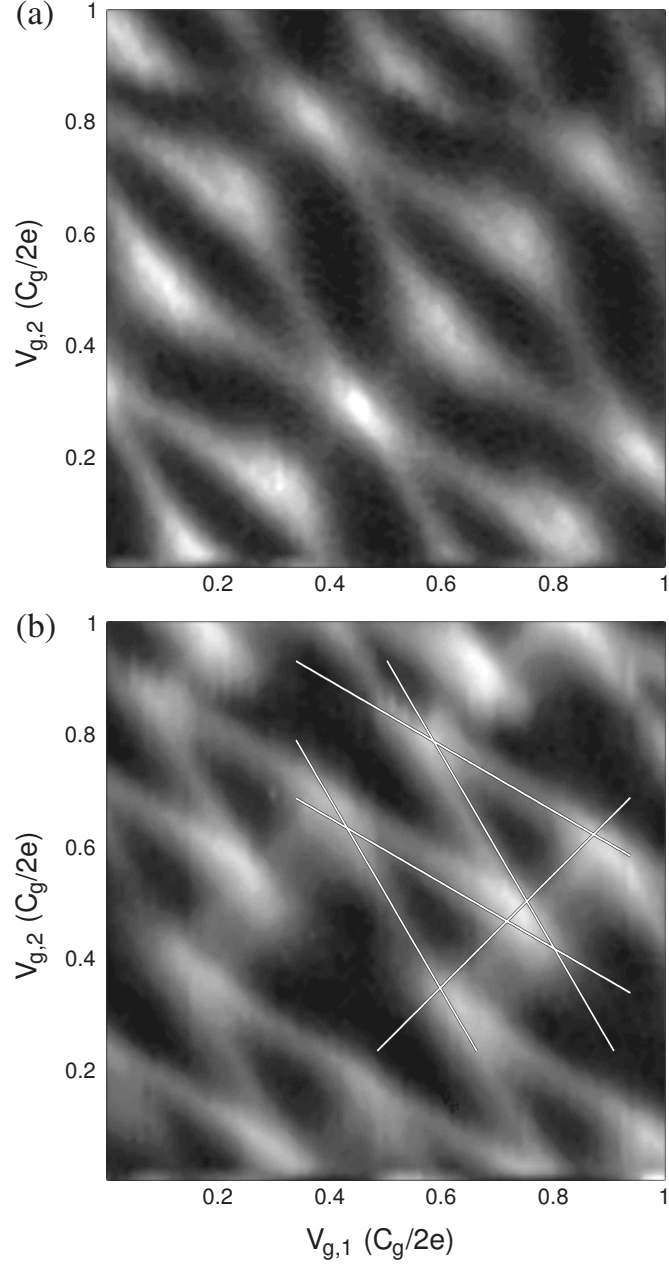}
\caption{Supercurrent $I_\mathrm{S}$ as a function of two gate voltages 
$V_{\mathrm{g},1}$ and $V_{\mathrm{g},2}$ with two different bias voltages 
applied: (a) $V\simeq 0$ (Adjusted to the maximum of the supercurrent) and
(b) $V=64$ $\mu$V. The lighter the color is, the higher is the current. In (b) 
the theoretically calculated resonance conditions for coherent tunnelling of 
Cooper pairs is presented as lines. These lines coincide with the pattern of
enhanced current in the experimental data.
\label{fig:Vhoney}}
\end{figure}

Applying bias voltage $V$ changes the
stability diagram so that each degeneracy node is split into a triangle
as discussed in Chapter \ref{sec:model}.
These triangles are clearly visible at the modulation plane 
measured at $V=64$ $\mu$V and shown in Fig.~\ref{fig:Vhoney}(b). 
Also theoretically calculated degeneracy lines giving a resonance 
condition for Cooper pair tunnelling, are drawn in the figure.
These lines are calculated using the parameters obtained earlier for 
the measured sample. Figure \ref{fig:Vhoney}(a) shows 
the same modulation plane but at $V\simeq 0$.
The triangles are formed around the nodes of each
honeycomb in Figs.~\ref{fig:Vhoney}(a) 
and \ref{fig:honey}, and the `extra' nodes formed of intersections
of different honeycombs disappear as they should. 

With a more detailed inspection one finds the measured 
current to be slightly increased  also at resonance lines for cotunnelling, 
which is not seen at the (current) scale of Fig.~\ref{fig:Vhoney}(b). 
Good agreement between theoretical resonance lines and
experimental data and also the reduction of the current inside the 
triangles proves the measured current to be mainly carried by the 
supercurrent, $I_\mathrm{S}$. Thus, quasiparticles 
present in the system are not acting as major carriers of current. 
The inelastic tunneling of Cooper pairs, where energy is interchanged
between the tunnelling Cooper pair and other parts of the system, e.g.,
electromagnetic environment or quasiparticles, is also not important.\cite{ing92}
Both these phenomena should increase current inside the whole 
triangle.\cite{bithesis}

The first resonance peaks in the I-V curve at $V\simeq 80$ $\mu$V 
(see inset in Fig.~\ref{fig:IV})
can also be explained by the effect of the bias to the stability diagram. 
At $V\simeq 0$ we have the highest current flowing at the triple nodes and
at the `extra' nodes, which are also three times degenerate. After 
increasing $V$ the nodes are split and the maximum current appears at 
the doubly degenerate sites, yielding much lower current than at
triple nodes of $V\!\sim\!0$. The new triply degenerate nodes are formed 
when the triangles situated close to each other overlap upon increasing 
$V$ and the maximum of the current is reached when vertices of 
the neighboring triangles coincide, as is approximately the case 
in Fig.~\ref{fig:Vhoney}(b). The closest nodes correspond to different 
quasiparticle configurations, $(0,0)$ and $(1,1)$ and are situated along the 
lines $V_{\mathrm{g},2} = V_{\mathrm{g},1} + k/2$, where $k$ is
an integer number, as seen in Figs.~\ref{fig:honey} and \ref{fig:Vhoney}(a).  
Using parameters obtained for this sample we estimate the 
new triple node to be formed between neighbouring zero bias nodes at
$V=83$ $\mu$V, which is exactly the voltage where the 
highest resonance peaks are located in the experimental I-V curve.

\section{Operation as a turnstile}

\subsection{Principle of operation}

A conventional turnstile for electrons or Cooper pairs consists of two 
arrays ($N\geq 2$) of junctions connected by a common island in between, 
whose charging and discharging can be controlled by a gate.\cite{and99,geer90c} 
The charging sequence is additionally controlled 
by the applied bias voltage, which determines the 
direction of the obtained DC-current when an AC-voltage 
is applied to the gate at frequency $f$. Each gate voltage cycle charges 
and discharges the island unidirectionally, thus generating a current $I=-ef$.

\begin{figure}[htb]
\linespread{1}
\includegraphics[width=90truemm]{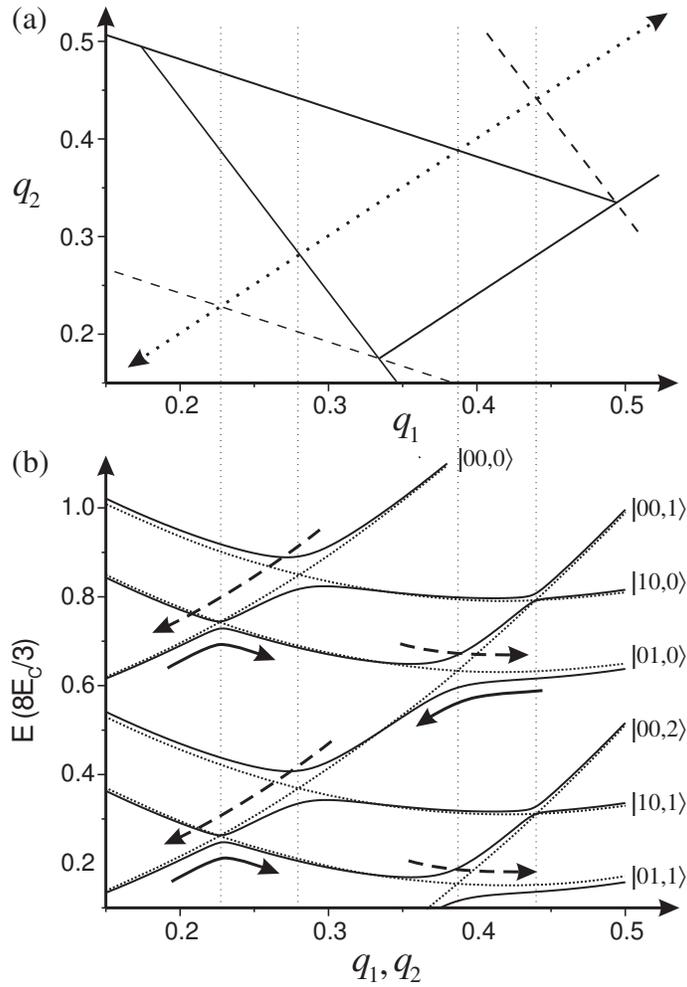}
\caption{(a) The path $[t\mapsto(q_1,q_2): q_1(t)=q_2(t)]$, 
used in the turnstile measurements (dotted line). It is centered 
at the degeneracy node of $V=0$ and exits the triangle around this node 
at both extremes. (b) Energy diagram of the CPP plotted along the path in (a). 
The solid lines represent the eigenstates of the system and the dashed lines 
are energies of the pure charge states of $H_\mathrm{Ch}$, which are
indicated at right by kets. The arrows show the method of transferring one 
Cooper pair per cycle through the CPP in the direction of the bias voltage as 
in a turnstile. The dashed arrows correspond to the relaxation into the lower
energy eigenstate at around resonances while the solid arrays indicate coherent 
tunnelling or cotunnelling. The thin vertical dotted lines are there to clarify
corresponding locations between (a) and (b). The diagram has been 
calculated using parameters $E_\mathrm{J}/E_\mathrm{C} = 0.2$ 
and $CV/2e = 0.16$. 
\label{fig:diagram}}
\end{figure}

The CPP can operate as a turnstile because of the hysteretic behaviour 
within a finite bias triangle, opened around node in the modulation 
plane. The simplest way to describe this behaviour is to consider 
a path in $(q_1,q_2)$ plane with the constraint 
$q_1\!=\!q_2$, exiting the triangle at both extremes (see Fig.~\ref{fig:diagram}(a)), 
and to assume that at every resonance, i.e., at each degeneracy of the charging 
Hamiltonian, $H_\mathrm{Ch}$, the system is driven to the state 
with lower energy. This simple reasoning alone is enough to explain the 
turnstile kind of behaviour: within every traversal of the path
one Cooper pair is transferred through the array in the direction of the 
bias voltage. This principle of operation involves coherent
tunnelling, cotunnelling and relaxation. It is pictured in 
Fig.~\ref{fig:diagram} and explained in more detail later. 

To obtain this hysteretic behavior one needs dissipation in the system. 
This can be provided, e.g., by the electromagnetic 
environment or quasiparticles, which both can absorb any amount of energy
dissipated in the system. Without dissipation 
the adiabatic passing through the degeneracy line
of charging energy $H_\mathrm{Ch} $ will retain the eigenstate of the system, 
i.e., the gap of $E_\mathrm{J}$ is not crossed. This will induce 
coherent tunneling of one Cooper pair in the direction which depends
whether the system is in the ground or in the excited state. 
The limit for the adiabaticity is determined by the probability of band crossing,
the so-called Landau-Zener (LZ) tunneling.\cite{ziman} This
probability can be written as\cite{zorset,hakosci}
\begin{equation}
P_\mathrm{LZ} = \exp\left(-\frac{\pi E_\mathrm{J}^2}
{8\hbar E_\mathrm{C}\dot q}\right) \equiv
\exp\left(-f_\mathrm{LZ}/ f\right),
\label{zener}
\end{equation}     
where $q=\sqrt{q_1^2 +q_2^2}$ is the absolute length in the $(q_1,q_2)$ plane.
In the absence of dissipation the probability for band crossing is 
equal in both directions, i.e., to excite or relax the system, but with 
dissipation the symmetry breaks. 
The probability to excite the system retains the same amplitude, 
$P_\mathrm{LZ}$, but the probability for relaxation
$P_\mathrm{Rel} = 1 - P_\mathrm{LZ} + P_\mathrm{LZ}^2$ 
is significantly increased and reaches almost unity
in the case of strong coupling to a dissipative element.\cite{ao,zorset}

How this effect of dissipation makes it possible to use CPP as a turnstile is 
explained in Fig.~\ref{fig:diagram}(b), where the energy diagram of the CPP is 
plotted along the path with time $t$ as a parameter 
$[t\mapsto(q_1,q_2): q_1(t)=q_2(t)]$ 
crossing the triangle as shown in Fig.~\ref{fig:diagram}(a). To be able 
to properly explain the behaviour we have to include the number
of Cooper pairs tunnelled through the array to our notation: so, instead of
$\vert n_1n_2\rangle$ we write $\vert n_1n_2, p\rangle$. 

If the system is initially
at the state $\vert 00,0\rangle$ and we start to increase both gates, i.e., moving 
from
left to right along the $x$-axis in Fig.~\ref{fig:diagram}(b), we come to the point 
of resonance for tunnelling through two junctions  simultaneously, i.e., cotunnelling,
to the state $|01,0\rangle$. This resonance condition is indicated by a dashed line in
Fig.~\ref{fig:diagram}(a). In the case of cotunnelling the total coupling 
for one Cooper pair to tunnel through two junctions, $E_\mathrm{J}^\mathrm{co}$, 
depends on the $E_\mathrm{J}$ of the junctions and the energy of the intermediate 
virtual state. Using the second order perturbation theory and the value $V=50$ 
$\mu$V used in measurements (see next section)  we obtain $E_\mathrm{J}^\mathrm{co}
/E_\mathrm{J} \approx 0.28$ yielding the Landau-Zener frequency  
$f_\mathrm{LZ}(E_\mathrm{J}^\mathrm{co}) \approx 65$ MHz.
This indicates the adiabatic condition to hold during gate excursion and cotunnelling 
to happen with frequencies smaller than 65 MHz.  
When continuing along the path, the system is retained in state $|01,0\rangle$
due to relaxation at resonances crossed. When coming back along the same path the 
system is driven to the state $|00,1\rangle$ due to a first order resonance, shown by
a solid line in Fig.~\ref{fig:diagram}(a), yielding coherent 
tunnelling and the system is kept there by relaxation (see Fig.~\ref{fig:diagram}(b)). 
Then the cycle starts over again with $\vert 00,1\rangle$ as an initial state.
Since the situation is fully asymmetric with respect to bias, the operation 
carries one Cooper pair per cycle through the array in the direction of the applied
bias voltage $V$.

\begin{figure}[htb]
\linespread{1}
\includegraphics[width=90truemm]{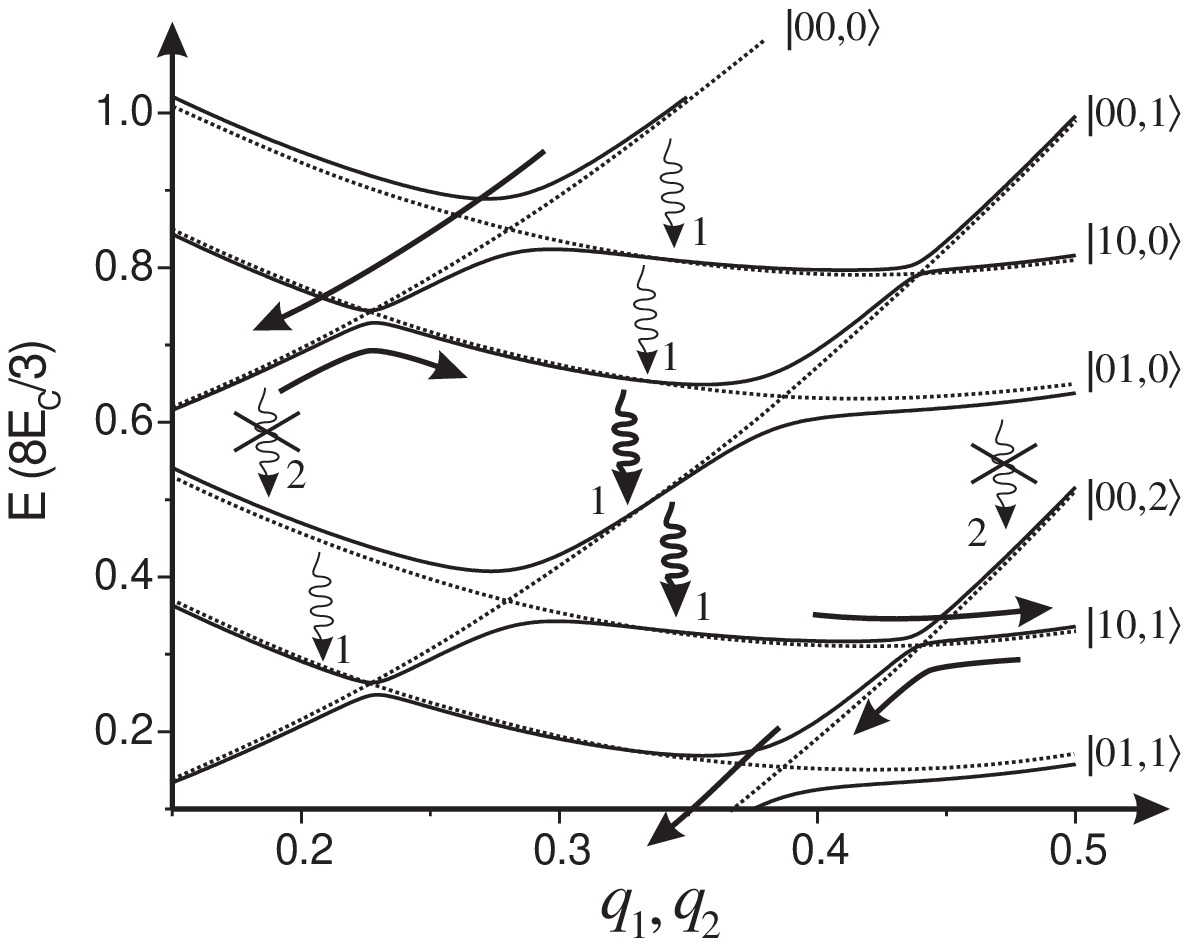}
\caption{Evolution of the CPP along the gate path of 
Fig.~\ref{fig:diagram}(a). The wavy arrows numbered as 1 show a couple 
of examples of possible undesirable inelastic tunnellings of 
a Cooper pair (or of two quasiparticles), which is the most significant 
source of errors in the cycle of Fig.~\ref{fig:diagram}(b). However, 
these are not affecting the final outcome of the gate cycle unless 
there happens several of them during one cycle.  
The thick solid (straight and wavy) arrows show an example of a cycle with
several inelastic tunnelling events. The charge transferred in that
particular cycle is two Cooper pairs, i.e., $4e$.
Striked out arrows numbered as 2 indicate similar inelastic 
relaxations which would instantly induce errors in pumping. These are fortunately 
suppressed due to the large number of intermediate tunnelling events needed.
\label{fig:diagram2}}
\end{figure}

The most significant source of errors in the cycle described above is the
inelastic tunnelling of a Cooper pair, which is also induced by the dissipation
and the probability of it is proportional to the function $P(E)$,\cite{ing92}
which gives the probability density of a tunnelling Cooper pair 
to emit (absorb) energy $E$ ($-E$) to the electromagnetic
environment. In addition, the presence of quasiparticles increases 
this probability by providing another way to exchange energy.
Examples of possible undesirable inelastic tunnelling events are shown
in Fig.~\ref{fig:diagram2} by wavy arrows numbered as $1$.
As one can figure out from the energy diagram, these are not
affecting the final outcome of the gate cycle unless there happens several
of them during one cycle. This would induce transfer of two or 
more Cooper pairs during that particular cycle as in the example
shown by thick solid arrows in Fig.~\ref{fig:diagram2}.
Also similar inelastic relaxation events which would instantly destroy the outcome
of the excursion, are indicated. Fortunately though, these relaxation
processes are likely to be largely suppressed due to the large number of virtual 
tunnelling
events, i.e., higher order of cotunnelling, needed in them. These processes are 
shown by thin wavy arrows striked out and numbered as $2$. Thus, the system is 
fairly rigidly 'locked' to transfer only one Cooper pair per cycle. 

\subsection{Measurements}

\begin{figure}[htb]
\linespread{1}
\includegraphics[width=85truemm]{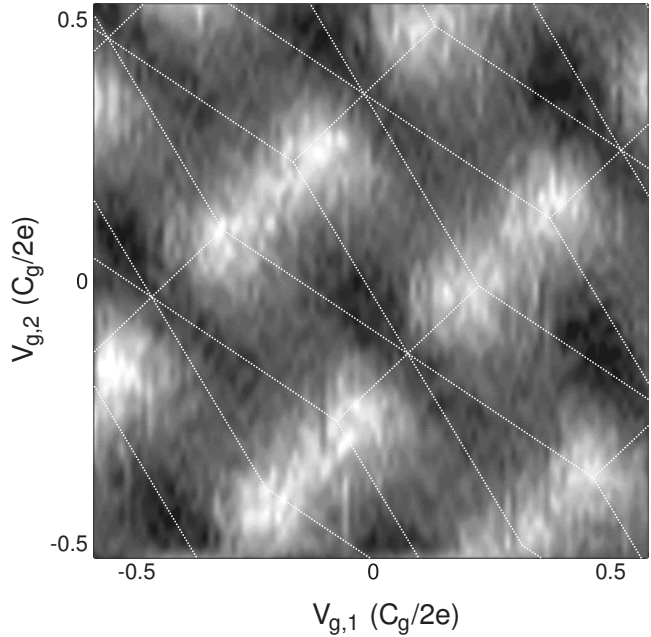}
\caption{Current with small bias voltage, $0\!<\!V\!<\!80$ $\mu$V, as 
a function of two DC voltages applied to gates $V_{\mathrm{g},1}$ and 
$V_{\mathrm{g},2}$ with 10 MHz in-phase sinusoidal signals added to
the two gates. The amplitude of 
the sine was $1/6\times 2e$-period. The measured structure of the modulation 
without RF signals and at $V \simeq 0$ is drawn as white lines.
\label{fig:TThoney}}
\end{figure}

To test whether the former line of reasoning holds we first measured the gate 
dependence of the current at a small bias voltage, $0\!<\!V\!<\!80$ $\mu$V, 
and 10 MHz sinusoidal signals
added on top of the DC voltages $V_{\mathrm{g},i}$ applied to 
both gates. The AC-signal was fed to both gates at the
same phase and the amplitude was $1/6$ times the $2e$-period, thus
corresponding to the path drawn in Fig.~\ref{fig:diagram}(a).
The obtained current as a function of DC gate voltages is 
plotted in Fig.~\ref{fig:TThoney}. The white dotted lines show
the structure measured without the RF signals at $V=0$. 
Data clearly indicate the enhancement of current inside the areas
of the triangles as predicted in the previous chapter.  

\begin{figure}[htb]
\linespread{1}
\includegraphics[width=90truemm]{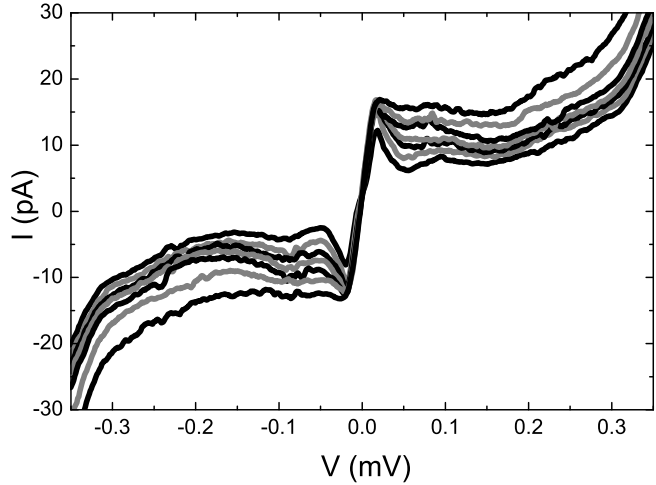}
\caption{The I-V curves measured with DC gate voltages tuned 
to one of the degeneracy nodes and in-phase sinusoidal RF signals added
to the two gates. Amplitude used was $1/6\times 
2e$-period and frequencies corresponding to I-V curves from lowest to
highest absolute current level are 0.5, 6, 12, 18, 26, 40 and 60 MHz. 
\label{fig:TTIV}}
\end{figure}

To obtain more quantitative results we froze the DC gate
voltages to the values corresponding to one of the degeneracy 
nodes and applied a similar sinusoidal signal as before
at different frequencies. The full  I-V dependence was measured 
instead of sitting at a fixed bias voltage to better
find the correct bias voltage, $0 \!<\! V \!<\! 83$ $\mu$V, where 
the applied gate path would be optimum for turnstile kind of behavior. 
Some of these measured I-V curves are shown in
Fig.~\ref{fig:TTIV}.

\begin{figure}[tb]
\linespread{1}
\includegraphics[width=90truemm]{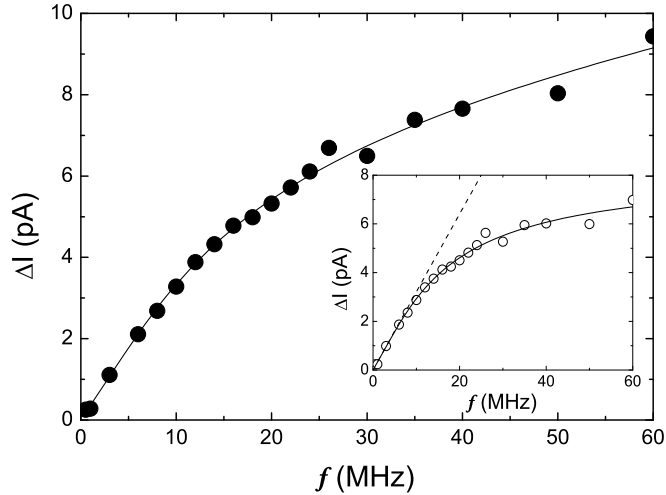}
\caption{Current in a cycle through the degeneracy node 
(see Fig.~\ref{fig:diagram}(a)) as a function of the frequency of the 
RF signals applied in-phase to the two gates. The bias voltage was $V=50$ 
$\mu$V and the AC-amplitude was $1/6\times 2e$-period. The solid line is the 
fit by Eq.~(\ref{eq:fitting}) and values for the fit parameters
are shown in the first line of Table \ref{table1} with the exception of $I_0$.
The inset shows the same current with the leak current subtracted 
and the dashed line shows the ideal $2ef$ dependence predicted by the theory. 
\label{fig:TT2ef}}
\end{figure}

From the I-V curves we can recognize the resonance point at $V\!\simeq\! 
80$ $\mu$V and the optimum operation point is the
minimum between that and zero voltage, i.e., around
$V\!\simeq\! \pm 50$ $\mu$V. If we extract the current at $V=50$ $\mu$V
and plot it against $f$ we obtain the dependence
shown in Fig.~\ref{fig:TT2ef}. The dashed line shows the ideal
$I\!=\!-2ef$ dependence which should be obtained if the device operates as a 
turnstile as explained in the preceding section. 

At low $f$ ($\leq 10$ MHz) the current increases as $I \!\simeq\! -2ef$ 
but starts to lack behind at around 25 MHz. To find out whether
this behaviour is consistent with LZ crossing, we fitted the data using
\begin{equation}
I = -2ef\left[1-P_\mathrm{LZ}(f)\right].
\end{equation}
The fitting parameters used were the starting slope ($\sim 2e$),
Landau-Zener frequency $f_\mathrm{LZ}$, and the offset in $I$.
This formula did not fit properly the data, mostly due to
the nonzero asymptotic slope of our data at high frequencies.  
To take this extra slope into account we modified the fitting 
function to allow for a finite leak current with linear dependence on $f$. 
The function which we used to fit the data and which is shown in 
Fig.~\ref{fig:TT2ef}, is of the form
\begin{equation} 
I = -2efQ_\mathrm{P}\left[1-\exp\left(-f_\mathrm{LZ}/f\right)\right] 
- 2efQ_\mathrm{L} + I_0,
\label{eq:fitting}
\end{equation}
where the fit parameters are $f_\mathrm{LZ}$, $Q_\mathrm{P}$ 
and $Q_\mathrm{L}$, which are the charges transferred and 
leaked during one cycle in units of $2e$, respectively, 
and $I_0$, which is the offset in current. 

The existence of this leak current can be physically justified as the result of
undesirable inelastic tunnelling of Cooper pairs. As explained earlier, these 
inelastic tunnelling events can happen inside the triangle during the operation 
and their contribution to the resulting current is, due to the fact that at the 
minimum two tunnelling events per cycle are needed to affect the 
outcome, small but nonzero. If we assume the 
probability for inelastic tunneling $P_\mathrm{IT}$ to be independent 
of frequency, we obtain an approximate expression for the leak current, 
$I_\mathrm{leak} = -[2eP_\mathrm{IT}^2 + 
\mathcal{O}(P_\mathrm{IT}^4)]f$, which depends 
linearly on $f$. The total current is thus of the form 
$I = -2ef[1-P_\mathrm{LZ}(f)]+I_\mathrm{leak}$ and 
$Q_\mathrm{L} \sim P_\mathrm{IT}^2$ in 
Eq.~(\ref{eq:fitting}).
 
Equation (\ref{eq:fitting}) fits perfectly the data and the values of the
fit parameters obtained using it are shown
in Table \ref{table1}. The analysis of the data was done both at 
negative and positive bias voltages. As explained earlier, the obtained
current changes sign with bias voltage. We also repeated the measurement 
in another node with slightly larger amplitude ($1/4\times 2e$-period) 
of RF signal and these results are also shown in Table \ref{table1}.

As seen in Table \ref{table1}, the obtained frequency 
dependence agrees with the theoretical prediction of $-2ef$ 
at low frequencies, i.e., $Q_\mathrm{P}\!\simeq\!1$. However, the Landau-Zener
frequency obtained, $f_\mathrm{LZ}\!\sim\!30$ MHz, is lower 
than that estimated ($\sim\,$50 MHz) for cotunnelling. 
This discrepancy can originate from  
the inhomogeneity of the sample. Since $P_\mathrm{LZ}$ is an 
exponential function of $E_\mathrm{J}$, it is the smallest $E_{\mathrm{J},i}$
in the array that determines the LZ threshold. The value $f_\mathrm{LZ}\!\sim\!30$ 
MHz yields $E_\mathrm{J}\!\approx\!17$ $\mu$eV which is a very reasonable value 
since the corresponding number estimated for a symmetric array is 19 $\mu$eV.
For the leaked charge we obtained $Q_\mathrm{L}\!\sim\!0.17$, which corresponds 
to a 17 \% probability for an extra Cooper pair to leak via inelastic tunnelling 
events
shown in Fig.~\ref{fig:diagram2}, during a cycle.

\begin{table}[htb]
\begin{tabular}{ccccc}
\hline\hline  
Ampl. (2e) & Bias & $Q_\mathrm{P}$ (2e)  & $f_\mathrm{LZ}$ (MHz)
& $Q_\mathrm{L}$ (2e)\\  \hline 
$1/6$ & $V\!>\!0\:$ & $0.985\pm 0.068 $&$ 26.2\pm 4.4 $&$ 0.127\pm 0.046 $\\ 
$1/6$ & $V\!<\!0\:$ & $0.956\pm 0.080 $&$ 29.1\pm 6.5 $&$ 0.142\pm 0.068 $\\
$1/4$ & $V\!>\!0\:$ & $0.996\pm 0.049 $&$ 35.1\pm 4.4 $&$ 0.132\pm 0.031 $\\
$1/4$ & $V\!<\!0\:$ & $0.976\pm 0.100 $&$ 24.7\pm 5.0 $&$ 0.229\pm 0.030 $\\ \hline 
$1/3$ & $V\!>\!0\:$ & $4.0\pm 2.3$ & $3.5\pm 1.8$ & $0.546\pm 0.029$ \\
$1/3$ & $V\!<\!0\:$ & $3.8\pm 1.3$ & $4.4\pm 1.4$ & $0.481\pm 0.027$ \\
\hline
\end{tabular}
\caption{Fit parameters using Eq.~(\ref{eq:fitting}) under various experimental
conditions. The current scale was shifted to set $I_0$ to zero.
\label{table1}}
\end{table}

From the preceding experimental results one can conclude that current in the in-phase 
gate cycles follows the relation $\Delta I\!=\!-2ef$ in data measured with amplitudes 
crossing only one triangle on the gate plane. However, this suggests that 
quasiparticle 
tunnelling plays a minor role when operating in this regime. This is contradicting the 
general 
assumption that quasiparticles are tunnelling completely stochastically in both time 
and direction, which was one of the explanations for the four honeycombs 
on the DC-modulation plane in section \ref{sec:mod}. This assumption of 
all stochastic behaviour should prevent the correct operation of our in-phase 
measurement as long as quasiparticle tunnelling is happening at the rate faster than 
the 
inverse measuring time $\tau_m^{-1}\!\leq\!(100$ $\mu$s$)^{-1}$, 
which indeed was assumed to explain the result of the DC-modulation measurement. 

If the quasiparticle tunnelling is happening 
less frequently than the frequency of the applied gate signals $f\!$, a major 
reduction in current should be observed due to missed cycles. Yet,
the whole principle of operation breaks down due to the undefined and 
constantly varying trajectory, if these tunnelling events are more 
frequent than $f\!$. Hence, neither of
these schemes yields $\Delta I=-2ef$.  

However, the relatively accurate behaviour in the in-phase measurements 
can be explained using the energy-minimisation model, which is also consistent
with the four honeycombs obtained in the DC-modulation measurements, as
explained earlier in section \ref{sec:mod}.
As long as the amplitude is small enough that the system
stays near one degeneracy node and thus at the largest possible current
given by the different choices of quasiparticle configurations, no quasiparticle 
tunnelling happens and the system is locked to a certain configuration yielding
accurate operation. But, as soon as we increase the amplitude too much 
the quasiparticle configuration starts to change and thus prevents the 
accurate transfer. However, if the gate trajectory only shortly goes 
out of the area of the locked configuration, quasiparticle tunnelling can
be prevented by the operating frequency $f$ being faster than
the time of quasiparticle tunnelling, and thus the configuration remains locked.

Increase in the bias voltage ($V\!\agt\!80$ $\mu$V) has the same effect 
as increasing the amplitude, since the high current areas (edges of the triangles) 
corresponding to different quasiparticle configurations start to overlap 
making the choice for the optimum quasiparticle configuration unclear. 
At the bias voltages higher than the optimum operation point $V\approx 50$ $\mu$V
the triangles of the stability diagram spread too much exceeding the amplitude 
used and the accuracy of the turnstile kind of behaviour is diminished until
at $V\!\agt\!80$ $\mu$V it breaks down. 
The effect of this can be seen in the I-V curves shown in Fig.~\ref{fig:TTIV},
which also show that some frequency dependence is still retained
after the resonance point $V\!\agt\!80$ $\mu$V, but we have no clear model
to explain this behaviour due to the existence of many overlapping hysteretic areas
corresponding to different quasiparticle configurations.  

\begin{figure}[bht]
\linespread{1}
\includegraphics[width=80truemm]{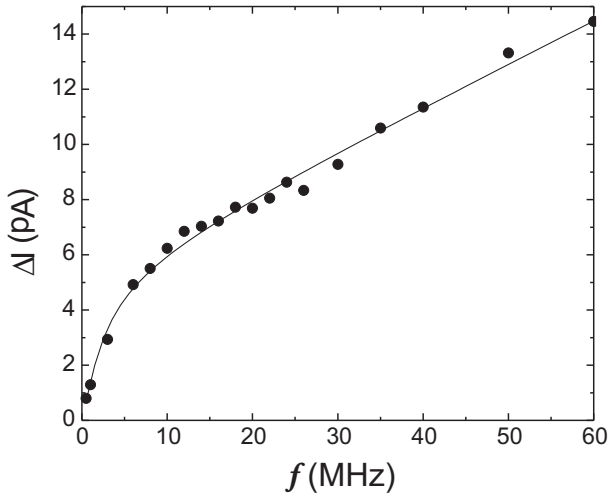}
\caption{Current at the degeneracy node as a function of 
the frequency of the applied in-phase RF signal of 
$\sim\!1/3\times 2e$-period amplitude. 
The circles correspond to data taken at the optimum bias voltage
$V\!\approx\!-50$ $\mu$V and the solid line is the fit by 
Eq.~(\ref{eq:fitting}) with parameters shown on the last 
row of Table \ref{table1}. 
\label{fig:TTlinear}}
\end{figure}

\begin{figure}[bt]
\linespread{1}
\includegraphics[width=80truemm]{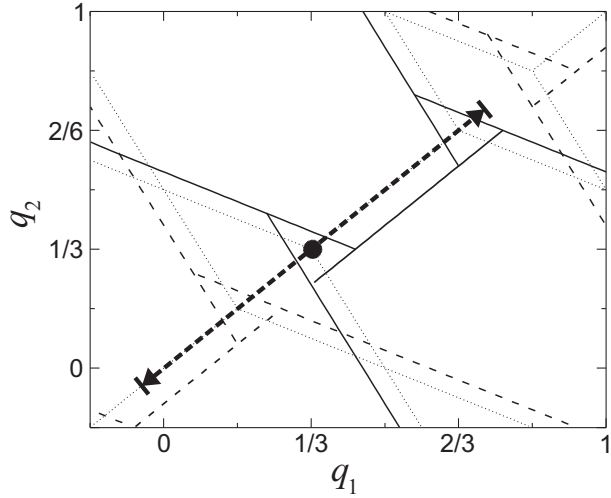}
\caption{Trajectory of gates, shown by the thick dashed arrow, in the in-phase 
measurement using the largest amplitude $\sim\!1/3\times 2e$-period. 
Thin dotted lines show two honeycomb patterns at $V\!=\!0$ corresponding to 
quasiparticle 
configurations $(0,0)$ and $(1,1)$. Black circle is the operating point 
tuned by DC gate voltages. Solid and dashed lines show the 
$2e$-periodic stability diagrams at $V\!=\!50$ $\mu$V corresponding to
quasiparticle configurations $(0,0)$ and $(1,1)$.
\label{fig:8e}}
\end{figure}

We also performed measurements using two times larger amplitude ($\sim\!1/3\times 
2e$-period) 
of the in-phase RF signal. The results obtained at $V=-50$ $\mu$V are shown in the 
last two lines of Table \ref{table1} and plotted in Fig.~\ref{fig:TTlinear} with 
a fit by Eq.~(\ref{eq:fitting}). In the absence of quasiparticle tunnelling and 
thus with a stable $2e$-periodic stability diagram, one would expect a 
twice larger $Q_\mathrm{P}$ ($Q_\mathrm{P}=2$) than in 
the data presented, e.g., in Fig.~\ref{fig:TT2ef}, 
since the trajectory of gates crosses two triangles during one cycle.
This trajectory is shown by the thick dashed arrow and the corresponding
$2e$-periodic stability diagram by solid lines in Fig.~\ref{fig:8e}. 

As expected, the measurements did not yield such a slope, but about 
$Q_\mathrm{P}\!\approx\!4$ instead, which can be explained with the 
energy-minimisation model. 
As seen in Fig.~\ref{fig:8e} we already move over to the neighbouring
triangle corresponding to a different quasiparticle configuration. Thus the
configuration is changed from (0,0) to (1,1) and back during every cycle.
These two extra quasiparticles responsible for changing the configuration are 
naturally
always driven in the direction of the bias voltage thus yielding extra charge of $2e$ 
carried during every cycle. If we take into account these two quasiparticles 
transferred during the cycle and the three triangles we cross, one corresponding
to quasiparticle configuration (1,1) and two to (0,0), as shown in 
Fig.~\ref{fig:8e}, we obtain $8e$ for the total charge carried per 
cycle. This explains the peculiar result of effectively four Cooper
pairs transferred per cycle. 

Measurement yields $f_\mathrm{LZ}\!\sim\!4$ MHz, which is lower than the one, 
$f_\mathrm{LZ}\!\sim\!14$ MHz, estimated with the smallest Josephson coupling 
of $E_\mathrm{J}\!\approx\!17$ $\mu$eV, obtained from the small amplitude 
measurements. 
This deviation is most likely due to the quasiparticle tunnelling involved in the 
cycle. 
Quasiparticle transfer has an upper limit of $\alt\,$5 
MHz,\cite{jen92,ave93,mart94,fon96} 
which together with the Landau-Zener limit of
Cooper pair transfer ($\sim\,$14 MHz) yields a smoother double transition starting 
from 
$\alt\,$5 MHz. The leak current is much higher compared to
the data taken with smaller amplitude and interestingly has the dependence of 
$I_\mathrm{leak}\!\approx\!-ef$ which might correspond to a leak of one quasiparticle 
during 
the cycle, but may as well be accidental and can also be explained by 50 \% 
probability of leakage of a Cooper pair during a cycle.

\section{Pumping measurements}

To complement the turnstile-type of in-phase measurements described in the previous 
chapter,
we also measured the same sample with 90$^\circ$ phase-shifted RF signals applied to 
the
two gates. This provided a circular path around the chosen degeneracy node. 
The node was again found as described earlier. At non-zero bias voltage one should 
traverse around the whole triangle to achieve proper pumping. 
Thus, nonzero bias voltage should allow pumping as long as neighbouring
triangles do not overlap. However, these measurements did not yield as 
clear results as the in-phase measurements. 

\begin{figure}[htb]
\linespread{1}
\includegraphics[width=80truemm]{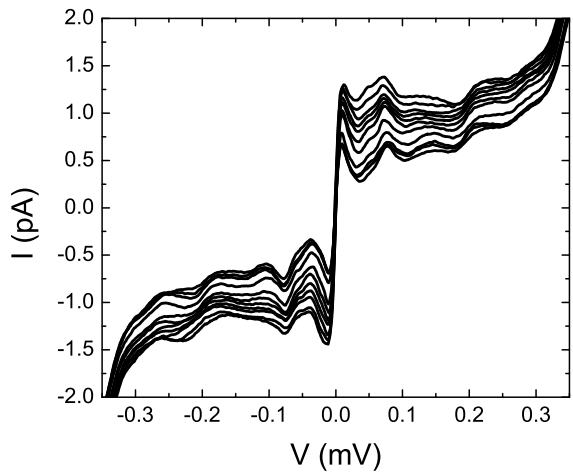}
\caption{I-V curves measured at one of the degeneracy nodes
with sinusoidal RF signal added to the two gates with 90$^\circ$
phase difference. The amplitude of the RF signal was $1/3\times
2e$-period and the frequencies were 0.1, 0.5, 1, 3, 6, 8, 10, 12, 14, 18, 
and 22 MHz from the smallest to the highest absolute level of current.
\label{fig:pumpiv}}
\end{figure}

We measured the pumped current at several different nodes and
with different amplitudes of the RF-signal: $1/9$, $1/6$, $1/4$ and $1/3$ times the 
$2e$-period. In Fig.~\ref{fig:pumpiv} some of the I-V curves measured using the 
largest of these amplitudes are shown. They exhibit clear frequency dependence which,
in turn, varies with the bias voltage. The most surprising detail in these I-V curves 
is 
that they reproduce the turnstile kind of behavior, so that the pumped current is 
always in the
direction of the bias voltage. Thus out-of-phase RF signals increase the bias driven 
current no matter what direction we wind around the node. 

\begin{figure}[htb]
\linespread{1}
\includegraphics[width=85truemm]{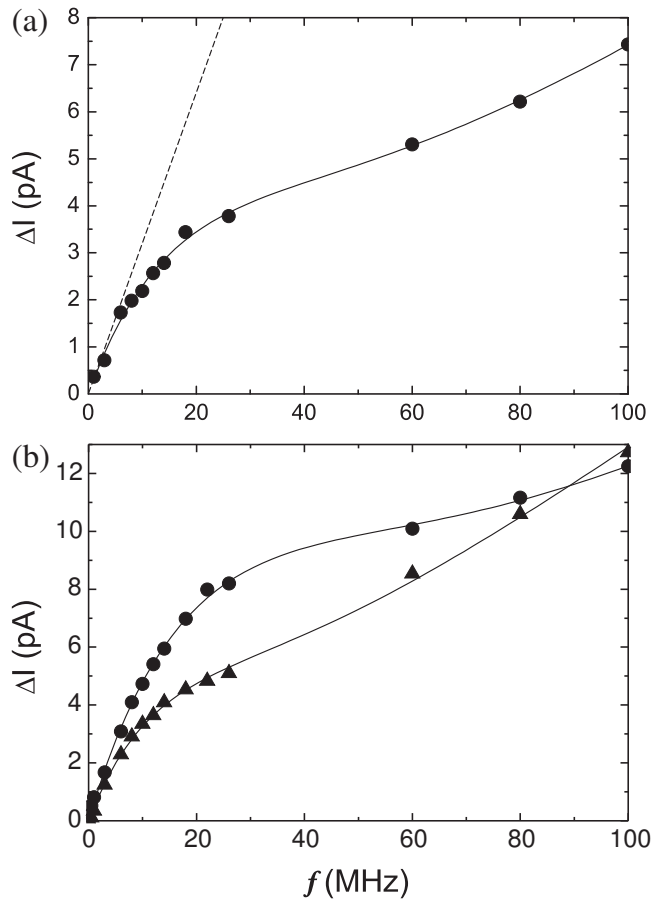}
\caption{Current as a function of the pumping frequency. The radius of
the circular trajectory of gates is $1/4\times 2e$-period in (a) and $1/3
\times 2e$-period in (b). The data in (a) have been taken at $V=365$
$\mu$V and  the dashed line corresponds to $\Delta I = -2ef$. Circles 
and triangles in (b) have been taken at $V=35$ $\mu$V ($<80 \mu$V)
and $V=305$ $\mu$V ($>80 \mu$V)
yielding the initial slope of $\sim 4e$ and $\sim 3e$, respectively.
Solid lines are fits by Eq.~(\ref{eq:2lines}), with parameters given in
Table \ref{table2}.
\label{fig:4ef}}
\end{figure}

To analyse the `pumped' current in more detail we examined the current 
at fixed voltages against frequency. Some examples of these plots are shown in 
Fig.~\ref{fig:4ef}. The general behavior of the current as a function of the 
frequency of the out-of-phase RF signals is similar to that observed in the in-phase 
measurements. The current increases first approximately linearly but the slope
decreases at around 30 MHz, which is again most likely due to Landau-Zener 
tunnelling. After that the behaviour has no general tendency. Curves corresponding
to different amplitudes or even to same amplitude but different bias  voltages 
differ a lot from each other. The only common feature is that the slope tends
to increase again. That is why the fitting of the formula similar to 
Eq.~(\ref{eq:fitting})
does not give a satisfactory result. 

To find the initial slope and an approximate 
value of the critical frequency $f_0\sim f_\mathrm{LZ}$ for 
LZ-crossover, we used an empirical function
\begin{equation}
I(f) \equiv -2efQ_\mathrm{P} + 2eQ_2\left[1 - 
exp\left(-f/f_0\right)\right]f + I_0,
\label{eq:2lines}
\end{equation}
which represents an exponential crossover from the initial slope 
$-2eQ_\mathrm{P}$ to another, $-2e(Q_\mathrm{P}-Q_2)$, with the 
transit frequency $f_0$. The fitting of this function gives more freedom at higher 
frequencies while it still reproduces the initial slope and $f_0$ with
good accuracy. Figure~\ref{fig:4ef} also shows the function $I(f)$ fitted
to the experimental data. The obtained slopes and critical frequencies from the 
fits using Eq.~(\ref{eq:2lines}) are shown in Table~\ref{table2}. 

The resonance point at $V\!\sim\!80$ $\mu$V, clearly visible also 
in Fig.~\ref{fig:pumpiv} is important in the analysis of these 
data as it was in the case of in-phase -measurements. 
At the two smallest amplitudes, $1/9$ and $1/6$ times $2e$-period,
frequency dependence of current was seen only at the resonance points. 
The slope obtained at this voltage is quite close to $2e\!\approx\!3.2\!\times\! 
10^{-19}$ A/Hz ($Q_\mathrm{P}\!\approx\!1$) with both amplitudes. 
At the amplitude of $1/4\times 2e$-period 
the plateau in current was spread over the whole bias range all the way up to the 
first JQP peak. At bias voltages smaller than the one corresponding to the resonance 
point the supercurrent masked the frequency dependence. Supercurrent is
very sensitive to the setting of gate voltages, and therefore current in this 
regime is susceptible to even small changes in the trajectory of gates.
At the bias voltages higher than the resonance point, systematic and rather stable 
frequency dependence was restored yielding the slope of 
$Q_\mathrm{P}\!\approx\!1$.  Data fits at
this RF-amplitude and bias voltage of $V\!=\!365$ $\mu$V are shown in 
Fig.~\ref{fig:4ef}(a).  The dashed line shows $\Delta I\!=\!-2ef$. At the highest 
amplitude, 
$1/3\times 2e$-period, the frequency dependence of current is very distinctive 
as seen in the I-V curves in Fig.~\ref{fig:pumpiv}. The obtained frequency 
dependencies split into two 
categories again. At bias voltages smaller than the resonance one $V\!\approx\!80$ 
$\mu$V the
initial slope is approximately $Q_\mathrm{P}\!\approx\! 2$ while at voltages
higher than the resonance one this slope is smaller with average at around 
$Q_\mathrm{P}\!\approx\! 1.5$. Examples of both of these 
cases are shown in Fig.~\ref{fig:4ef}(b).  The obtained critical frequency, 
$f_\mathrm{0}$, is always of the same order $\sim 30$ MHz as in the in-phase
measurements, thus further suggesting it to originate from Landau-Zener events. 

\begin{table}[htb]
\begin{tabular}{ccccc}
\hline\hline 
Ampl. (2e) & V ($\mu$V) & $Q_\mathrm{P}$(2e) &~~& $f_\mathrm{0}$ (MHz) \\ \hline 
$1/9$ & $-83$ &$0.65\pm 0.12$ & &$48\pm 16$\\
$1/9$ & $83$ &$0.99\pm 0.20$ & &$31.5\pm 7.0$\\ \hline 
$1/5$ & $-83$ & $0.97\pm 0.18$ & &$26.4\pm 4.6$\\
$1/5$ & $83$ & $0.97\pm 0.34$ & &$26.2\pm 9.0$\\ \hline
$1/4$ & $250\rightarrow 350$ & $1.00\pm 0.26$ & &$24.0\pm 7.0$\\
$1/4$ & $-350\rightarrow-70$ & $0.886\pm 0.060$& &$29.0\pm 2.4$\\  
$1/4$ & $365$ & $1.000\pm 0.074$ & &$22.1\pm 1.6$ \\
$1/4$ & $-320$ & $0.934\pm 0.086$ & &$23.1\pm 2.3$ \\
$1/4$ & $310$ & $1.05\pm 0.15$ & &$22.9\pm 3.4$ \\ \hline
$1/3$ & $75$ & $1.91\pm 0.10$ & &$25.1\pm 1.7$ \\
$1/3$ & $55$ & $1.986\pm 0.062$ & &$28.2\pm 1.2$ \\
$1/3$ & $35$ & $2.130\pm 0.060$ & &$32.4\pm 1.4$ \\
$1/3$ & $-75$ & $1.529\pm 0.094$ & &$24.1\pm 1.9$ \\
$1/3$ & $305$ & $1.53\pm 0.10$ & &$16.6\pm 1.1$ \\
\hline
\end{tabular}
\caption{Results of the data fits of the out-of-phase measurements. 
Parameters as in Eq.~(\ref{eq:2lines}).  
\label{table2}}
\end{table}

It is noteworthy that out-of-phase gate control does not make a 
distinction whether the phase difference is $+90^\circ$ or $-90^\circ$.
This is the second feature besides the bias dependence indicating that we 
do not pump charges in the traditional sense. Direction of the obtained
current is determined only by the bias voltage whereas the pumping frequency
and radius determine the magnitude. This can be explained in 
approximately the same way as the in-phase measurement, where
the system always chose the lowest energy state when 
passing a resonance. When trying to pump against the bias 
voltage the resonance point for cotunnelling to happen is reached before the 
resonance of single tunnelling as can be seen in Fig.~\ref{fig:triangle}.
It is likely that in our experiment cotunnelling is so strong that, 
it always happens first. This cycle of cotunnelling 
yields a transfer of two Cooper pairs during one cycle, which is twice as much 
as in pumping in the direction of bias.\cite{geer} However, this was not the case 
in our measurements which yielded $|\Delta I| = 2ef$ in both directions. This
discrepancy could be due to $\sim\,$33 \% suppression of cotunnelling, which 
sounds reasonable as we can estimate $f_\mathrm{LZ}$ for cotunnelling in 
case of, e.g., the amplitude of $1/4\times 2e$-period, to
be of the order of few MHz, which would suggest rather high suppression.  

One possible explanation for the discovered behaviour arises from the circular 
pumping trajectory. Due to the structure of the stability diagram, 
$E_\mathrm{J}^\mathrm{co}$ between the states $\vert 10\rangle$ 
and $\vert 01\rangle$ along the circular path is suppressed by the factor 
$\sqrt{5/2}$ more as compared to other cotunnelling events along the 
trajectory. This directly leads to the reduction of $f_\mathrm{LZ}$ to 
$2/5\times f_\mathrm{LZ}$ and thus to almost doubling of $P_\mathrm{LZ}$
 in the middle junction at $f\approx f_\mathrm{LZ}\sim 2$ MHz.
This again could prevent the cotunnelling in the middle junction (but not in the other 
two) and 
thus yield a $\sim\,$33 \% suppression of the total cotunnelling probability, 
which would prevent a cycle transferring one Cooper pair against the bias voltage 
as in the experiments.

\section{Discussion}

To further discuss the validity of the energy-minimisation model
and the model developed for the turnstile-kind of behaviour, we briefly
review some of the earlier measurements of the CPP or similar devices.
Somewhat similar turnstile-kind of behaviour in the case of a superconducting 
SET has been reported earlier  in Ref.~\onlinecite{zorset}. The principle
of operation in that particular experiment was based on a similar asymmetric
behaviour of the system when passing a resonance point.\cite{ao}
The main difference besides the number of junctions (two in that case), was 
that in that experiment Cr resistors were embedded near the sample to 
provide the required dissipation. These resistors also suppressed
any current near the zero voltage and thus the current plateaus
were only seen at a finite voltage between $E_\mathrm{C}/2e$ and
$2\Delta/e$. (Our measurement indicates that no additional 
resistances are necessary to obtain this behaviour.)
Relatively accurate result of the experiment is consistent with the 
energy-minimisation model, since the gate trajectory they used in
Ref.~\onlinecite{zorset} stayed all the time close to a resonance 
point and thus close to the peak in supercurrent. 
However, this peak in supercurrent was suppressed due to the
resistive environment, which might have increased the 
probability of quasiparticle tunnelling, yielding a reduced
accuracy in `pumping': This agrees with the experiment as they
reported current which was inferior to $I\!=\!-2ef$ at the plateaus observed. 

Our out-of-phase measurement along the circular path around the
degeneracy node also yielded current $-2ef$ 
but the direction was defined only by the bias voltage as in the in-phase
measurements. The proper direction of current, depending on the 
direction of pumping and not on the bias voltage, 
was observed in the measurements of the Cooper 
pair pump by Geerligs {\it et al.}\cite{geer} This difference 
is most probably due to the smaller ratio $E_\mathrm{J}/
E_\mathrm{C}\approx 0.03$ in that experiment, 
which reduced the probability of cotunnelling significantly. 
This suppression is also consistent with a lower Landau-Zener frequency, 
$f_\mathrm{LZ}$, obtained in that experiment and the lack of the 
zero bias supercurrent, which prevented the measurement of the
DC-modulation. Yet, a small step in a pumped current at $V=0$ 
remained as a consequence of cotunnelling, but the small height of it
indicates much higher suppression of cotunnelling than in our experiment.
Geerligs {\it et al}. also state that the step was significantly larger 
in another sample, even reversing the current, which is what we observe.

The pumping trajectory used in Ref.~\onlinecite{geer} was very wide
and according to the energy-minimisation model the quasiparticle configuration 
should have changed several times during each cycle. On the contrary
the measurement yielded rather accurate behaviour, which would indicate
no change of the quasiparticle configuration during the operation. This 
could be explained by the fact that quasiparticles had no freedom to tunnel and 
a certain configuration remained locked all the time, thus resulting in pure 
$2e$-periodic stability diagram, which is also reported,
e.g., in Ref.~\onlinecite{levy02}. The failure of measuring the DC modulation 
in the superconducting state in Ref.~\onlinecite{geer} can be explained by 
the lack of supercurrent (small $E_\mathrm{J}$), since there should not 
be current flowing inside the opened triangles of the 
stability diagram at $V\neq 0$, which 
was used to map the degeneracy nodes in the normal state.
This can be verified by the measurement of Fig.~\ref{fig:Vhoney}(b).

In the measurement of Ref.~\onlinecite{zorpump} a similar idea of
resistive environment as in Ref.~\onlinecite{zorset} was used, how in the
case of a CPP. The influence of the resistive environment was to suppress
the otherwise strong cotunnelling of Cooper pairs. The experimental results
showed suppression of current near the zero bias as in 
Ref.~\onlinecite{zorset} and thus the current plateaus were again obtained
only at higher bias voltages. These plateaus were very weak and yielded 
a behaviour similar to our measurement. The value of the pumped current
was approximately $-2ef$ but the direction was determined just by the bias voltage. 
This could be due to inability of the Cr resistors to prevent the cotunnelling
in spite of the small ratio $E_\mathrm{J}/E_\mathrm{C}\approx 0.013$. 
As stated in Ref.~\onlinecite{zorpump} the quasiparticles 
were actively present also in this experiment.

\section{Conclusions}

We have shown that current in both the in-phase and out-of-phase gate cycles 
follows the relation $\Delta I \!=\! -2ef$ in data measured with amplitudes 
crossing or encircling only one triangle on the gate plane. These experiments also 
demonstrate in practise how Cooper pair pump could be used as a turnstile with help of 
dissipation. A model was developed to explain the process and experiments clearly 
demonstrated this behavior quantitatively.

The failure to observe pure $2e$-periodicity in the DC modulation
measurement indicates an active presence
of non-equilibrium quasiparticles in the system, as in many earlier 
experiments.\cite{geer,geer90b,zorpump,laf91} However, the agreement 
between the model developed for the turnstile-kind of behaviour
and the measurements suggests that quasiparticle tunnelling has a minor 
effect when operating in this regime, which is contradicting the general 
assumption that quasiparticles are tunnelling completely stochastically. To explain 
this twofold behavior of the system, a model was developed, which is 
based on a very general tendency of a system to always strive for minimum
energy. It suggests that if the quasiparticles in the system have freedom to 
move, they will organize themselves to the configuration giving the highest 
current, which continuously lowers the energy of the system. The four 
honeycombs obtained in the DC-modulation measurement are consistent
with this model and the relatively accurate behavior in the RF-measurements 
can be explained with it, too. The model also agrees with the earlier measurements
of the Cooper pair pump\cite{geer,zorpump} and another similar device.\cite{zorset} 

We also measured the current pumped through the array by winding around a 
degeneracy node along a circular path. This showed the correct magnitude
in obtained current but the direction was not defined by the direction 
of pumping but it was rather determined by the bias voltage thus
reproducing the turnstile kind of behavior. This behaviour could be 
explained in general terms, although not as precisely as in the case 
of in-phase measurements, by the similar energy minimisation argument, 
that the system chooses the state 
corresponding to a lower energy when the resonance is passed adiabatically.  
This does not fully explain the observed magnitude of current $|I|=2ef$
when pumping in the direction against the bias voltage. But, it agrees with a strong 
cotunnelling in our experiment, which is due to the large ratio of
$E_\mathrm{J}/E_\mathrm{C}\approx 0.15$.

As a final note we state that it is very unlikely for Cooper pair pump
as such to be able to provide a current standard or otherwise work with 
high accuracy. The strong cotunnelling and relaxation, among other
uncontrollable processes, tend to degrade the pumping cycles. 
However, there 
might be ways, such as embedding the sample in a highly resistive 
environment\cite{zorpump} or using a combined flux and charge 
control,\cite{sluice,rom02} to overcome these difficulties. 
Also the use of the CPP to measure, e.g., decoherence
time would need much more controlled electromagnetic 
environment to be successful.\cite{pek01,faz03} 

\begin{acknowledgments}
The authors wish to thank A.~Halvari for fabrication 
of the sample and K.~Hansen and M.~Aunola for fruitful 
discussions and comments. This work has been supported by the Academy of Finland
under the Finnish Centre of Excellence Programme 2000-2005
(Project No. 44875, Nuclear and Condensed Matter Programme at JYFL)
and partially supported by EU (Project SQUBIT2), Finnish Graduate School 
of Material science and V\"ais\"al\"a foundation.
\end{acknowledgments}


\begin{thebibliography}{60}
\bibitem{sampo} The circuit has 
been designed and constructed by Kari Loberg and 
the controlling software has been done using LabView$^{\textcircled{\scriptsize
R}}$ by Sampo Tuukkanen at the Department of Physics, University of Jyv\"askyl\"a.
\end{thebibliography}

\end{document}